\documentclass[twocolumn,aps,pra,showpacs,preprintnumbers, superscriptaddress, amsmath,amssymb]{revtex4}

\usepackage{stmaryrd} 
\usepackage{mathtools} 
\usepackage{stmaryrd} 

\usepackage{amsmath}
\usepackage{amsfonts}

\usepackage{amssymb}
\usepackage{graphicx}%

\begin{document}

\author{Andre G. Campos}
\email{agontijo@princeton.edu}
\affiliation{Department of Chemistry, Princeton University, Princeton, NJ 08544, USA}

\author{Renan Cabrera}
\affiliation{Department of Chemistry, Princeton University, Princeton, NJ 08544, USA} 

\author{Denys I. Bondar}
\affiliation{Department of Chemistry, Princeton University, Princeton, NJ 08544, USA} 

\author{Herschel A. Rabitz}
\affiliation{Department of Chemistry, Princeton University, Princeton, NJ 08544, USA} 

\title{ Violation of Hudson's theorem in relativistic quantum mechanics}

\date{\today}

\begin{abstract}
 In non-relativistic quantum mechanics, Hudson's theorem states that a Gaussian wave-function is the only pure state corresponding to a positive Wigner function (WF). We explicitly construct non-Gaussian Dirac spinors with positive relativistic WF. These pure relativistic states are coherent superpositions of particles and antiparticles, while the existence of positive WF exclusive composed of particles is conjectured. These observations may open new directions in relativistic quantum information theory.
\end{abstract}

\pacs{03.65.Pm, 05.60.Gg, 05.20.Dd, 52.65.Ff, 03.50.Kk}

\maketitle

\emph{Introduction.} In non-relativistic quantum mechanics the phase-space representation of the density 
operator is known as the Wigner quasi-probability distribution \cite{Wigner1932,PhysRevA.59.971,Hillery1984121,Polkovnikov2010}, which is broadly applied in studying
the quantum-to-classical transition \cite{blokhintsev2010philosophy, Heller1976, zurek2001sub, zachos2005quantum, Kapral2006,zurek1991decoherence}, optics, signal processing \cite{dragoman}, and quantum computing \cite{emerson, mari}. The Wigner function is not an ordinary phase-space distribution as it is often negative. Moreover, the Wigner function's negativity is a resource for quantum computation speed up, whereas strictly positive Wigner functions can be efficiently simulated by classical algorithms \cite{emerson, mari}. 

In an influential paper, Hudson \cite{Hudson1974249} demonstrated that, in non-relativistic quantum mechanics, the only pure state with a positive Wigner function is a Gaussian. In this Letter we show that there are non-Gaussian relativistic pure states corresponding to positive Wigner functions. Moreover, the explicitly constructed states are shown to be a superposition of positive energy (usually interpreted as particles) as well as negative energy (interpreted as antiparticles) solutions of the Dirac equation. Furthermore, the time evolution of a free particle does not preserve positivity of the Wigner function, while the evolution of a coherent superposition of the first two Landau states in the presence of a homogenous magnetic field preserves positivity of the Wigner function. Due to a vital role of the Wigner's function negativity in quantum information processing \cite{emerson, mari}, these unique features of relativistic systems open new opportunities in the emerging field of relativistic quantum computation \cite{Martin-Martinez}.

\emph{Relativistic quantum mechanics in the phase space.} The celebrated Dirac equation reads
\begin{align}
 \left[c \gamma^0 \gamma^{\mu} (i \hbar \hat{\partial}_{\mu}-\frac{e}{c}A_{\mu})  -  \gamma^0 m c^2 \right] \psi(x) = 0,
    \label{standardDirac}
\end{align}
where $\gamma^0=\beta$ and $\gamma^k=\beta\alpha^k$ with $\alpha^k$ and $\beta$ being the Dirac matrices and repeated indices are summed over $\mu = 0,1,2,3$ and $k=1,2,3$.

The relativistic matrix-valued Wigner function for spin $1/2$ particles in explicitly covariant form was put forth in Refs. \cite{Hakim2011introduction,elze1986transport,hakim1982,hakim1978covariant,vasak1987quantum}, while a non-explicitly covariant form in quantum field theory was developed in Ref. \cite{Bialynicki-Birula1991}. The non-explicitly covariant form of the relativistic Wigner function is employed throughout because of its convenient direct comparison with the non-relativistic counterpart.

A relativistic extension of the phase-space formalism leads to the following Wigner matrix 
\begin{align}
 W(t,x,p) =  \frac{1}{(2\pi)^3}\int d^3 \theta \, B(t,x,\theta) \exp( i p\cdot \theta  ),
\label{J-Wigner}
\end{align}
where
\begin{align}
  B(t,x,\theta) \equiv  \psi\left( t,x -\frac{\hbar}{2} \theta  \right) \psi^{\dagger}\left( t,x + \frac{\hbar}{2} \theta\right)\gamma^0,
  \label{B}
\end{align}
with $\psi$ being a solution of the Dirac equation (\ref{standardDirac}). The zeroth component of the Wigner matrix (\ref{J-Wigner}), defined as
\begin{align}
   W^0(t,x,p)={\rm Tr}\, [W(t,x,p)\gamma^0]/4,
   \label{wigner}
 \end{align}
 realizes a phase-space representation of the Dirac spinor $\psi$, namely the marginals $\int W^0(t,x,p) dp$ and $\int W^0(t,x,p) dx$ coincide with the coordinate and momentum probability distributions, respectively. Moreover, the expectation value of an observable $\hat{G} = G(\hat{x},\hat{p})$ obeys
\begin{align}
\langle \hat{G} \rangle= \langle \psi | \hat{G} | \psi \rangle = \int dx dp W^0(x,p) G(x,p),
\end{align}
where $G(x,p)=\int d\theta \langle x-\hbar\theta/2|G(\hat{x},\hat{p})|x+\hbar\theta/2\rangle e^{ip\theta}$. Therefore, we shall refer to $W^0(t,x,p)$ as the \textit{relativistic Wigner function}. Further details about the phase-space representation of relativistic quantum mechanics can be found in Refs. \cite{bolivar2001classical,vasak1987quantum,Bialynicki-birula1977,Shin1993,Hakim2011introduction}. 

\emph{States with positive relativistic Wigner functions.} 
First, consider the spinor
 \begin{eqnarray}
\psi=\frac{\mathcal{C}e^{-\frac{x^2}{2\sigma^2}}}{\sqrt{2mc(mc+p^0)}}\begin{pmatrix}
         p^0+mc\\
         0  \\
         i(p^0+mc)  \\
         0   
     \end{pmatrix}, \,
              \label{matrix}
\end{eqnarray} 
where $\mathcal{C}$ is a normalization constant. The state (\ref{matrix}) corresponds to a strictly positive relativistic Wigner function,
\begin{eqnarray}
	W^0(x,p)&=&\frac{1}{(2\pi)^3}\int d^3 \theta\sum_{n=1}^{4}\psi_n(x_L)\psi_n^\dagger(x_R) e^{ i p_{k} \theta^{k}  }\nonumber\\
&=&\frac{\mathcal{N}}{(2\pi)^3}\int d^3 \theta  e^{ i p_{k} \theta^{k}  }e^{-\frac{(x_L)^2}{2\sigma^2}}e^{-\frac{(x_R)^2}{2\sigma^2}}\nonumber\\
&=&2\sqrt{\pi}|\sigma|\mathcal{N}e^{-\frac{x^2+\sigma^4p^2}{\sigma^2}},
	\label{proof}
\end{eqnarray}
where $\mathcal{N}$ is a positive constant, $x_L=x -\hbar \theta/2$, and $x_R=x + \hbar \theta / 2$. The following non-Gaussian modification of the state (\ref{matrix})
 \begin{align}
 \psi(x)=\mathcal{C}e^{-a^2x^2}\begin{pmatrix}
         1\\
         0  \\
         ax  \\
         0   
     \end{pmatrix},
 \label{first_excited}
 \end{align}
corresponds to the positive relativistic Wigner function (Fig. \ref{fig5}):
 \begin{align}
W^0(x,p)=\frac{1}{2a^2\pi}\mathcal{C}e^{-\frac{p^2}{2a^2} - 2a^2x^2}\left(p^2  + 4a^4x^2\right).
\label{magnetic}
\end{align}
Note that the states (\ref{matrix}) and (\ref{first_excited}) are the ground and first excited states for an electron in a  homogenous magnetic field \cite{dirac}, respectively.

Additionally, the state
\begin{align}
\psi(x)=\mathcal{C}\begin{pmatrix}
         e^{-(-b+x)^2/a^2}+e^{-(b+x)^2/a^2}\\
         0  \\
         e^{-(-b+x)^2/a^2}-e^{-(b+x)^2/a^2}  \\
         0   
     \end{pmatrix},
     \label{positive_a}
 \end{align}
has a strictly positive relativistic Wigner function  
\begin{align}
W^0(x,p)=4\sqrt{2\pi}|a|\mathcal{C}e^{-\frac{a^2p^2}{2}-\frac{2}{a^2}(b^2+x^2)}\cosh\left(\frac{4bx}{a^2}\right),
\label{wigner_1}
\end{align}
 for any values of real parameters $a$ and $b$. The following generalization of the state (\ref{positive_a})
 \begin{align}
 \psi(x)=\mathcal{C}\begin{pmatrix}
         qe^{-(-d+x)^2/a^2}+(1-q)e^{-(b+x)^2/a^2}\\
         0  \\
         qe^{-(x-d)^2/a^2}-(1-q)e^{-(b+x)^2/a^2}  \\
         0   
     \end{pmatrix},
 \label{positive_b}
 \end{align}
has also a positive Wigner distribution
\begin{align}
W^0(x,p)=2\sqrt{2\pi}\mathcal{C}|a|\left(e^{-\frac{a^4p^2+4(b+x)^2}{2a^2}}(q-1)^2+\right. \nonumber\\
+\left. q^2e^{-\frac{a^4p^2+4(d-x)^2}{2a^2}}\right),
\label{wigner_2}
\end{align}
 for any real values of $q$, $b$, $a$, and $d$. 
 
\begin{figure}
  \includegraphics[scale=0.25]{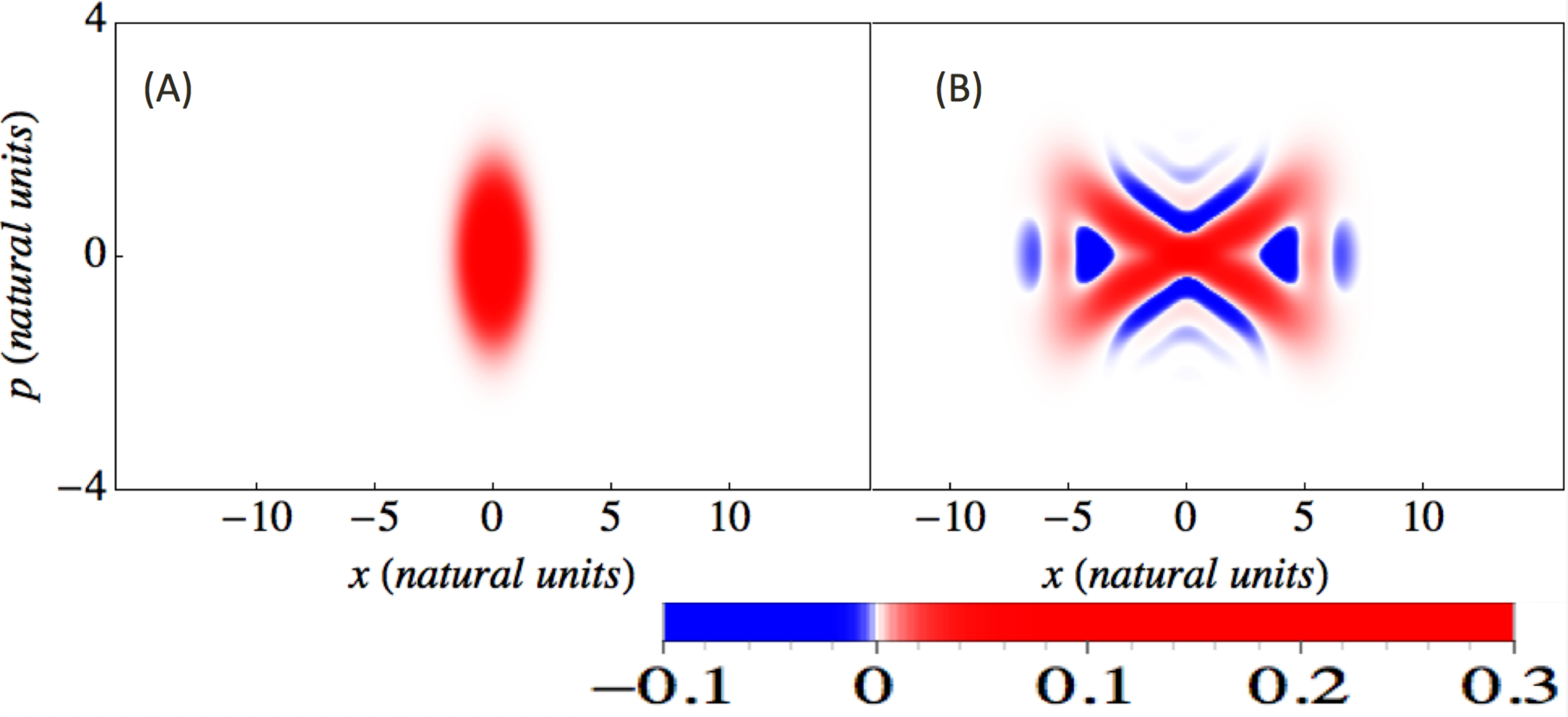}
  \caption{ (color online) (A) The relativistic Wigner function of the spinor (\ref{matrix}) at time $t=0$ in natural units. (B) The relativistic Wigner function of the time-propagate state at $t=7.7$ in natural units. Blue and red colors denote negative and positive values, correspondingly.} 
     \label{fig1}
\end{figure}

\begin{figure}
  \includegraphics[scale=0.25]{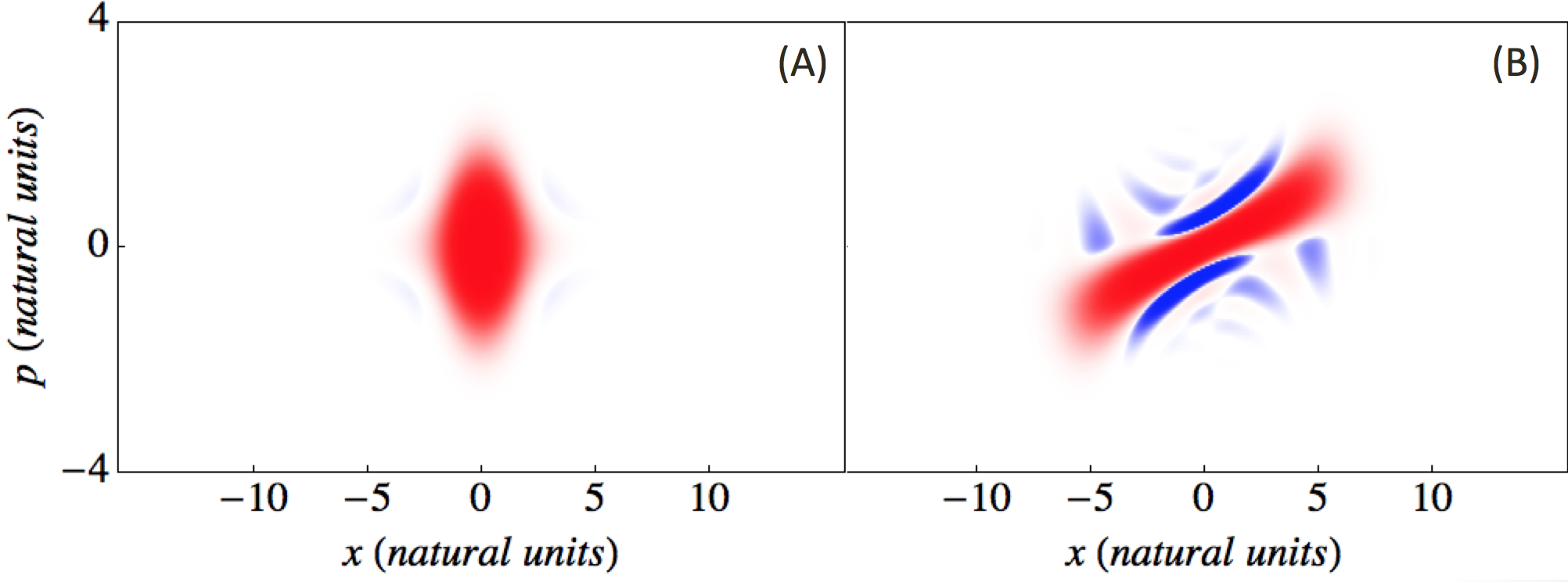}
  \caption{ (color online) (A) The relativistic Wigner function after filtering out antiparticles in the state (\ref{matrix}) at time $t=0$ in natural units.  (B) The relativistic Wigner function of the time-propagate state at $t=7.7$ in natural units. Blue and red colors denote negative and positive values, correspondingly.}
     \label{fig2}
\end{figure}

\begin{figure}
  \includegraphics[scale=0.25]{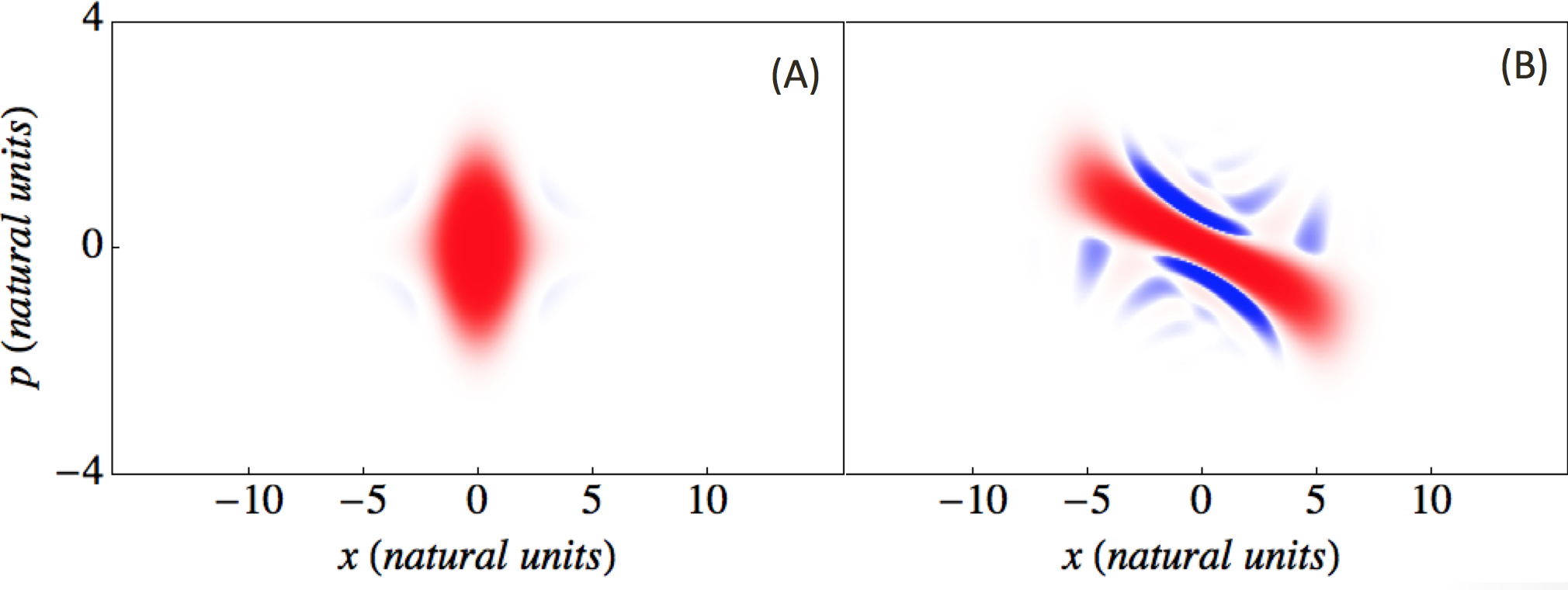}
  \caption{ (color online) (A) The relativistic Wigner function after filtering out particles in the state (\ref{matrix}) at time $t=0$ in natural units.  (B) The relativistic Wigner function of the time-propagate state at $t=7.7$ in natural units. Blue and red colors denote negative and positive values, correspondingly.}
     \label{fig3}
\end{figure}

\emph{Illustrations.} 
Employing the numerical method for the single-particle Dirac equation from Ref. \cite{Fillion-Gourdeau2011}, we establish that the states (\ref{matrix}), (\ref{positive_a}), and (\ref{positive_b}) are coherent superpositions of positive (i.e., particles) and negative (i.e., antiparticles) energy solutions \cite{greiner2000relativistic}. In particular, time-propagation of the initial state (\ref{matrix}) by the free Dirac equation is shown in Fig. \ref{fig1}, where we see the formation of an x-shaped packet at a later time due to \textit{zitterbewegung} \cite{greiner2000relativistic, thaller,thaller1} -- the interference between particles and antiparticles. The origin of the `x' shape is elucidated by dynamics depicted in Figs. \ref{fig2} and \ref{fig3}. If the antiparticles (particles) are projected out \cite{greiner2000relativistic} from the initial state (\ref{matrix}), then subsequent time propagation by the Dirac equation forms the first (second) half of the x-shaped wave packet in Fig. \ref{fig1}(B).

The phase-space dynamics of the state in Fig. \ref{fig2}, exclusively composed of particles, is consistent with non-relativistic dynamics:  Positive momentum wave packets move in positive directions. However, the dynamics of antiparticles (Fig. \ref{fig3}) is a mirror image of the dynamics in Fig. \ref{fig2}: Positive momentum wave packets move in negative directions. The latter is a direct consequence of the CPT theorem \cite{greiner2000relativistic}, stating that the Dirac equation is invariant under simultaneous charge conjugation C (which replaces particles by antiparticles and \emph{vice versa}), parity transformation P, and time reversal T.

According to Figs. \ref{fig1}-\ref{fig3}, the negativity of the relativistic Wigner function (blue color) is not preserved during the free particle evolution of the Dirac equation. This even holds for wave packets initially consisting of only particles (Fig. \ref{fig2}) or antiparticles (Fig. \ref{fig3}). This situation is significantly different from non-relativistic  free particle evolution, which always conserves the Wigner function's negativity \cite{moyal}. Moreover,  contrary to the non-relativistic case, a superposition of states (\ref{proof}) and (\ref{first_excited}) maintains the positivity of the Wigner function during propagation in a homogeneous magnetic field, complying with the fact that magnetic fields block pair creation \cite{PhysRevA.86.013422}.

 \begin{figure}
  \includegraphics[scale=0.25]{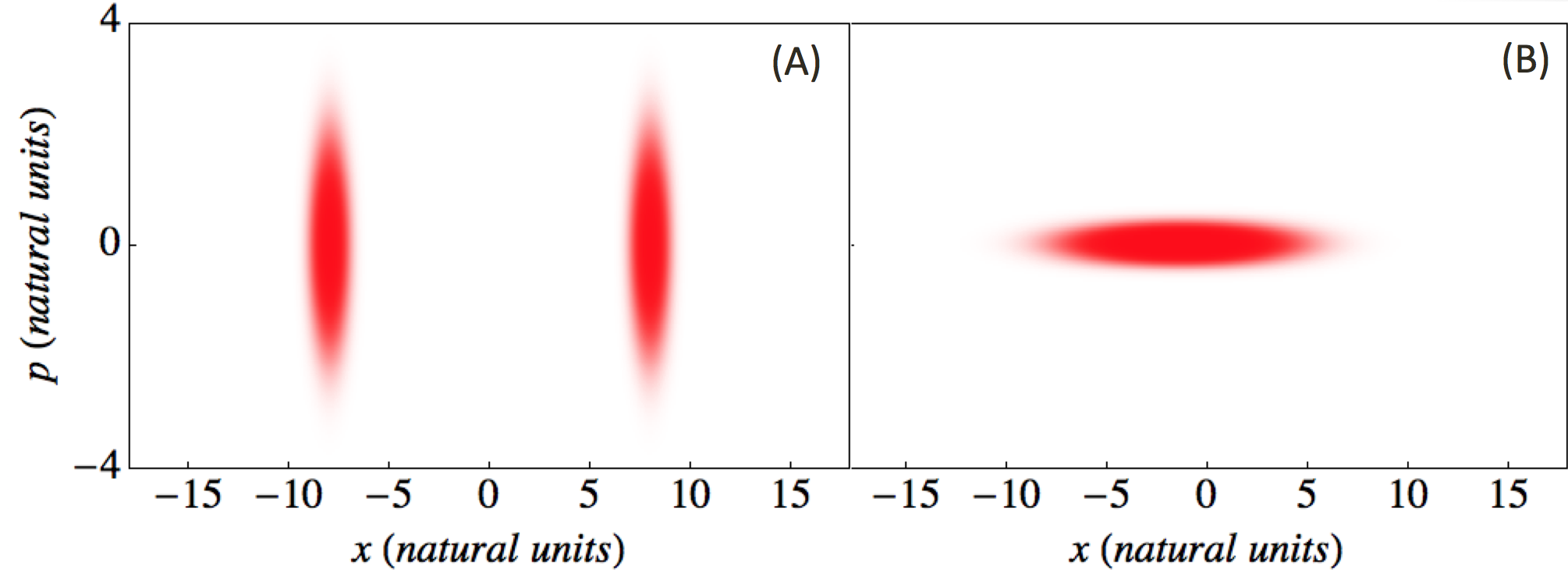}
  \caption{(color online) (A) The positive Wigner function for the spinor (\ref{positive_a}) with $a=\sqrt{2}$ and $b=3$ in natural units. (B) The Wigner function for the spinor (\ref{positive_b}) with $q=0.1$, $a=5$, $b=3$, and $d=0.1$ (in natural units) after removal of antiparticles.}
     \label{fig4}
 \end{figure} 
  \begin{figure}
  \includegraphics[scale=0.25]{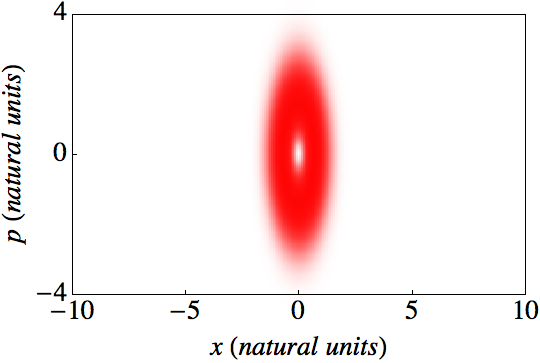}
  \caption{(color online) The positive Wigner function for the non-Gaussian spinor in Eq. (\ref{first_excited}) for $a=1$, which can be associated to the first excited state of the Dirac equation for an electron in a constant and homogenous magnetic field.}
     \label{fig5}
 \end{figure} 

Particles or antiparticles can be projected out directly in the Wigner picture without involving the Dirac spinors. Utilizing the Hilbert phase space approach \cite{Bondar2012, Bondar2013}, we can obtain the following projector in the phase space
\begin{equation}
\mathcal{P}_s^{\pm}=\frac{1}{2}\left(\textbf{1}+s\frac{c(p_k\pm\hbar\lambda_k/2)\alpha^k+\beta mc^2}{\sqrt{(p_k\pm\hbar\lambda_k/2)^2c^2+m^2c^4}}\right),
  \label{filter}
\end{equation}
where the operators $x_k$, $p_k$, and $\lambda_k$ obey $[x_k, \lambda_l] = i\delta_{k,l}$, $[x_k, p_l] = [\lambda_k, p_l] = 0$, $\textbf{1}$ is a $4\times4$ identity matrix, and $s=1$ ($s=-1$) is chosen to filter antiparticles (particles) out. Then, the filtering of an arbitrary state is accomplished in two steps
\begin{eqnarray}\label{Ws_projected}
C_s(t,\lambda,p)&=&\frac{1}{(2\pi)^6}\int\int d^3\theta d^3x \, \mathcal{P}^+_{s}B(t,x,\theta)\nonumber\\
&&\times\mathcal{P}^-_{s}\exp(ip\cdot\theta)\exp(-ix\cdot\lambda)\nonumber\\
W_s(t,x,p)&=&\frac{1}{(2\pi)^3}\int d^3\lambda C_s(t,\lambda,p)\exp(i\lambda\cdot x), 
\end{eqnarray}
where $W_s(t,x,p)$ is the resulting Wigner distribution and $B(t,x,\theta)$ is defined in Eq. (\ref{J-Wigner}).
Note that Eqs. (\ref{filter}) and (\ref{Ws_projected}) are applicable to pure as well as mixed relativistic states.

Figure \ref{fig4}(a), depicting the Wigner function for the state (\ref{positive_a}), reveals another unique feature: The relativistic Wigner function of a coherent superposition of two spatially non-overlaping wave packets need not display interference, which always shows up in the non-relativistic phase space \cite{Wigner}. To clarify this observation, consider a family of strictly positive Wigner functions $W^0_{\psi_n}(x,p)$ for the spinors $\psi_n(x)$. For $\chi(x)=a\psi_1(x)+b\psi_2(x)$, one obtains
\begin{align}
W^0_{\chi}(x,p)=&|a|^2W^0_{\psi_1}(x,p)+|b|^2W^0_{\psi_2}(x,p)+\nonumber\\
&+ a^\ast bW^0_{\psi_1^\dagger\psi_2}(x,p)+b^\ast aW^0_{\psi_2^\dagger\psi_1}(x,p).
\label{wigner4}
\end{align}
The relativistic Wigner function (\ref{wigner4}) remains strictly positive as long as $\psi_1(x)$ is orthogonal to $\psi_2(x)$. For example, the spinor (\ref{positive_a}) can be represented as the superposition of the following spinors
 \begin{eqnarray}
\psi_1=e^{-(x-b)^2/a^2}\begin{pmatrix}
         1\\
         0  \\
         1  \\
         0   
     \end{pmatrix},\quad
\psi_2=e^{-(x+b)^2/a^2}\begin{pmatrix}
         1\\
         0  \\
         -1  \\
         0   
     \end{pmatrix}.
     \label{spinor}
  \end{eqnarray}
such that $W^0_{\psi_1^\dagger\psi_2}(x,p)=W^0_{\psi_2^\dagger\psi_1}(x,p)=0$, leading to a strictly positive phase space distribution. On the other hand, the superposition of 
 \begin{eqnarray}
\psi_1=e^{-(x-6)^2/2}\begin{pmatrix}
         1\\
         0  \\
         i  \\
         0   
     \end{pmatrix},\quad
\psi_2=e^{-(x+6)^2/2}\begin{pmatrix}
         1\\
         0  \\
         i  \\
         0   
     \end{pmatrix},
     \label{spinor}
  \end{eqnarray}
displays a significant interference as typically observed in the non-relativistic phase space \cite{interference}.

Note that  according to Eq. (\ref{proof}), the relativistic Wigner function for a spinor with components $\phi_n(x)$ is mathematically analogous to the non-relativistic Wigner function for the density matrix $\rho = \sum_{n=1}^4 \phi_n(x) \phi_n^*(x)$. This correspondence not only is responsible for the violation of Hudson's theorem in relativity, but also allows to adapt non-relativistic bounds on Wigner's function positivity \cite{mixed_wigner}.

Given the discussion so far one may ask, are there Dirac spinors with positive Wigner functions made only of particles? We have not found a conclusive answer to this question. However, the following numerical evidence allows us to conjecture the existence of such states: Figure \ref{fig4}(B) depicts the numerical Wigner transform of the state resulting from projecting out antiparticles from spinor (\ref{positive_b}) with $q=0.1$, $a=5$, $b=3$, and $d=0.1$. The area of negative values in Figure \ref{fig4}(B) are of the order of round-off errors ($\sim 10^{-15}$).
  
\emph{Outlook.} 
Contrary to non-relativistic mechanics, there exist whole families [e.g., see Eqs. (\ref{first_excited}), (\ref{positive_a}) and (\ref{positive_b})] of pure non-Gaussian states underlying positive relativistic Wigner functions (\ref{wigner}). Additionally, a superposition of the first two (but not higher) Landau levels [Eqs. (\ref{proof}) and (\ref{first_excited})]  has a strictly positive Wigner function at any time during evolution in a homogeneous magnetic field. Further departing from the non-relativistic picture, the explicitly constructed states are composed of both particles and antiparticles. It is also shown that the free Dirac evolution does not preserve the volume of negativity in the phase space. 

We additionally developed the procedure of filtering particles or antiparticles out directly in the phase space [Eq. (\ref{Ws_projected})], which can be applied not only to pure states (i.e., what a majority of other methods perform) but also to an arbitrary mixed state. This procedure further extends the techniques currently utilized in relativistic quantum chemistry \cite{liu1,liu2}.

\emph{Acknowledgments} The authors acknowledge financial support from NSF CHE 1058644, DOE DE-FG02-02-ER-15344 and ONR-MURI W911-NF-11-1-2068. A.G.C was also supported by the Fulbright foundation.

\bibliography{bib-relativity}

\end{document}